\documentclass[review]{elsarticle}

\usepackage{lineno,hyperref}

\usepackage{graphicx}
\usepackage{dcolumn}
\usepackage{bm}
\usepackage{rotating}
\usepackage{xcolor}

\usepackage{tikz}
\usepackage{graphicx}
\usepackage{chngcntr}
\usepackage{mathtools}
\usepackage{amsfonts}

\usepackage{subfigure}

\DeclareUnicodeCharacter{03B3}{$\gamma$}
\usepackage{textgreek}
\usepackage{multirow}
\usepackage{booktabs,siunitx,makecell}

\modulolinenumbers[5]

\journal{Journal of \LaTeX\ Templates}

\begin{document}

\begin{frontmatter}

\title{Thickness Dependent Sensitivity of GAGG:Ce Scintillation detectors for Thermal Neutrons: GEANT4 Simulations and Experimental Measurements}

\author[mymainaddress,mysecondaryaddress,mydepartment]{Annesha Karmakar}


\author[mysecondaryaddress]{G. Anil Kumar\corref{mycorrespondingauthor}}
\cortext[mycorrespondingauthor]{Corresponding author}
\ead{anil.gourishetty@ph.iitr.ac.in}

\author[Barcaddress]{Mohit Tyagi}
\author[mydepartment]{Anikesh Pal}

\address[mymainaddress]{Nuclear Engineering and Technology Program, Indian Institute of Technology Kanpur, India}
\address[mysecondaryaddress]{Radiation Detectors and Spectroscopy Laboratory, Department of Physics, Indian Institute of Technology Roorkee, India}
\address[Barcaddress]{Technical Physics Division, Bhabha Atomic Research Centre, Mumbai, India}
\address[mydepartment]{Department of Mechanical Engineering, Indian Institute of Technology Kanpur, India}

\begin{abstract}
In the present work, we report extensive GEANT4 simulations in order to study the dependence of sensitivity of GAGG:Ce scintillation crystal based detector on thickness of the crystal. All the simulations are made considering a thermalised Am-Be neutron source. The simulations are validated, qualitatively and quantitatively, by comparing the simulated energy spectra and sensitivity values with those obtained from experimental measurements carried out using two different thicknesses of the crystal from our own experiment (0.5mm and 3mm)  and validated with three other thicknesses (0.01mm, 0.1 mm and 1 mm) from literature \cite{Taggart2020}. In this study, we define sensitivity of GAGG:Ce as the ratio of area under 77 keV sum peak to 45 keV peak. The present studies clearly confirm that, while it requires about 0.1 mm thickness for the GAGG:Ce crystal to fully absorb thermal neutrons, it requires about 3 mm to fully absorb the thermal neutron induced events. Further, we propose an equation, that can be used to estimate the thickness of the GAGG:Ce crystal directly from the observed sensitivity of the GAGG:Ce crystal. This equation could be very useful for the neutron imaging community for medical and space applications, as well as for manufactures of cameras meant for nuclear security purposes.

\end{abstract}

\begin{keyword}
\texttt  \quad GAGG:Ce \sep GEANT4 \sep Sensitivity  \sep Energy Response \sep Nuclear imaging 
\end{keyword}

\end{frontmatter}

\section{Introduction}
Developments involving neutron detection are being pursued in various fields like homeland security, nuclear safeguards, nuclear decommissioning radioactive waste management, space application and nuclear medical uses. Thermal neutron detection, in particular has been seen to rely heavily on Helium-3 and Boron Trifluoride gas-filled tube based detectors \cite{Kojima2004}. However, with the dwindling supply of Helium-3 gas and toxic nature of the gaseous medium, researchers are working on best alternatives to these gas-filled detectors such as  scintillation detectors based on Gadolinium, Lithium and Boron due to their high thermal neutron capture cross sections \citep{Kojima2004,VanEijk2012,Cieslakcieslak2019}. Among these, GAGG:Ce proved to be promising candidate for thermal neutron detection due to its largest thermal neutron capture cross section ($^{157}Gd$ - 257000 barns and $^{155}Gd$ - 60900 barns), non-requirement of expensive isotropic enrichment, fast decay time, good compatibility of emission band with SiPMs, non-hygroscopic in nature and lack of internal radioactivity. It also possesses high effective atomic number $(Z=55)$, good light yield $(~55000 ph/MeV)$, good gamma energy resolution $(6\% \quad at \quad  662 keV)$ and high density $(6.7 g/cm^3)$. These attractive properties are responsible for its applications in the fields of medical imaging, homeland security, high energy gamma measurements, etc. \citep{Kato2013, Rawat2018EfficiencySO, Tyagi2020}. Several studies were reported in the literature on the response of GAGG:Ce scintillators for thermal neutrons. The absorption of thermal neutrons causes $^{155}Gd$ and $^{157}Gd$ isotopes to enter their excited states. During de-excitation, gamma rays having energies up to around 8.5 MeV are produced \cite{Xu2013}. In addition, conversion electrons of energies around 45 keV and X-rays of energies around 32 keV are also produced. Whenever these conversion electrons and X-rays deposit their energies within the resolving time of the detector, a photo-peak around 77 keV appears in the spectrum due to the effect of true coincidence summing \citep{Tyagi2019,Dumazert2018}.  Naturally, the simulations of the response of GAGG:Ce scintillation detectors for thermal neutrons are of great importance, as simulations can help in understanding the effect of various parameters related to the radiation source and detector on the detector performance. These parameters include source-detector geometry, type of radiation, energy of the radiation, thickness of detector, etc. Recently, Rawat et al., have studied the effect of source-detector geometry and energy on the absolute efficiencies (both total detection and full-energy absorption) of GAGG:Ce detectors for gamma rays \cite{RawatS}. In the present work, the effect of GAGG:Ce crystal thickness on the sensitivity for thermal neutrons is studied.\\
The GAGG:Ce detector was shown to exhibit good performance, where the thickness was one of the main features \cite{Nikl2006,Kanai2013,Sen2017,Gerasymov2020}. While the GAGG:Ce crystal should be thin enough to absorb thermal neutrons, there should also be sufficient thickness for complete X-ray absorption in order to provide a high spatial resolution. Since Gadolinium has an exceptionally high thermal neutron absorption cross-section, the detector becomes an excellent choice for thermal neutron detection \cite{Dumazert2018,Taggart2019,Tyagi2019,Tyagi2020} but its main inherent drawback is its increased gamma-ray sensitivity, eg, natural radioactive background. A study reported neutron absorbing capabilities of Gadolinium loaded garnets (GAGG and GYAGG) \cite{Dosovitskiy2019}. While the study could establish the capability of detecting neutrons from an Am-Be source exceeding 1 eV, the characteristic lines of neutron detection were not reported for both experimental and GEANT4 studies. The neutron detection capability of GAGG:Ce detectors was also presented by Korjik et.al \cite{Korjik2019}. The study mainly focused on the neutron detection capability which was analysed using GEANT4 and later experimentally verified using Am-Be source. However, the characteristic peaks around ~45keV and ~77keV for thermal neutron response of GAGG:Ce was not reported, rather a peak of 500 keV was reported as a signature for thermal neutron detection. There have been few studies reported on the thickness dependence on the sensitivity of GAGG:Ce detectors for neutron detection. Fedorov et.al \cite{Fedorov2020} defined the sensitivity of GAGG:Ce detector as the number of counts per second per unit of neutron flux in $(cm^2/s)^{-1}$. It was found that for GAGG:Ce in thermal neutron detection mode, 10mm thickness is excess and can be significantly reduced without critical loss of sensitivity of neutrons, rather an improvement in neutron/gamma-ray ratio. However, the signature peaks of GAGG:Ce detector for thermal neutron detection for both GEANT4 simulation and experiment did not quite agree and only the 77 keV was reported. Taggart et.al \cite{Taggart2020} reported that the key to optimization of GAGG:Ce thickness for neutron detection is the balance between sufficient material to stop the incident neutron and subsequent de-excitation while limiting sensitivity to background high energy gamma rays that deposits energy within the scintillator. The performance metric is the ratio of the amplitude of the ~82keV peak to its background coined as a signal-to-background ratio. This signal-to-background ratio improves significantly with reduced thickness. However, due to the limitation of available scintillators, the maximum signal-to-background ratio could not be established. Additionally, no simulation studies are reported in order to estimate the optimum sensitivity of GAGG:Ce detector for thermal neutron detection.\\
The present work aims to study the response of GAGG:Ce scintillation crystal based detector for thermal neutrons in order to understand the dependence of the sensitivity on the crystal thickness. Since the thermal neutron induced events shows up in form of two signature peaks, 45 keV and 77 keV, where the latter is dominant,  we define sensitivity as the ratio of 77 keV sum peak to 45 keV peak. To the best of our knowledge, for the first time we present detailed study of GAGG:Ce using GEANT4 simulation toolkit considering a thermalised Am-Be source \citep{Allison2006}. Simulations were done in the following steps: 1) Modelling of the Am-Be neutron source, 2) Determining the moderator thickness to achieve maximum thermal neutron flux, 3) Establishment of physics list suitable for simulating thermal neutron response of GAGG:Ce, 4) Mean free path of GAGG:Ce for thermal neutrons, 5) Establishing an equation for correlating thickness and sensitivity of GAGG:Ce for thermal neutrons. In order to validate the simulations on thickness dependent sensitivity, experimental measurements were done considering two different thicknesses, namely, 0.5mm and 3mm of the GAGG:Ce crystal and a thermalised Am-Be source. The simulated results were also compared with measurements reported in the literature for three different thicknesses, namely, 0.01mm, 0.1 mm and 1 mm \cite{Taggart2020}. The results so obtained were used to establish a quantitative relation between the sensitivity of GAGG:Ce and crystal thickness so that crystal thickness can be determined directly from the thermal neutron induced spectrum of GAGG:Ce scintillation crystal based detector.

\begin{figure}[tbt]
 \centering
 \includegraphics[height = 2.5in]{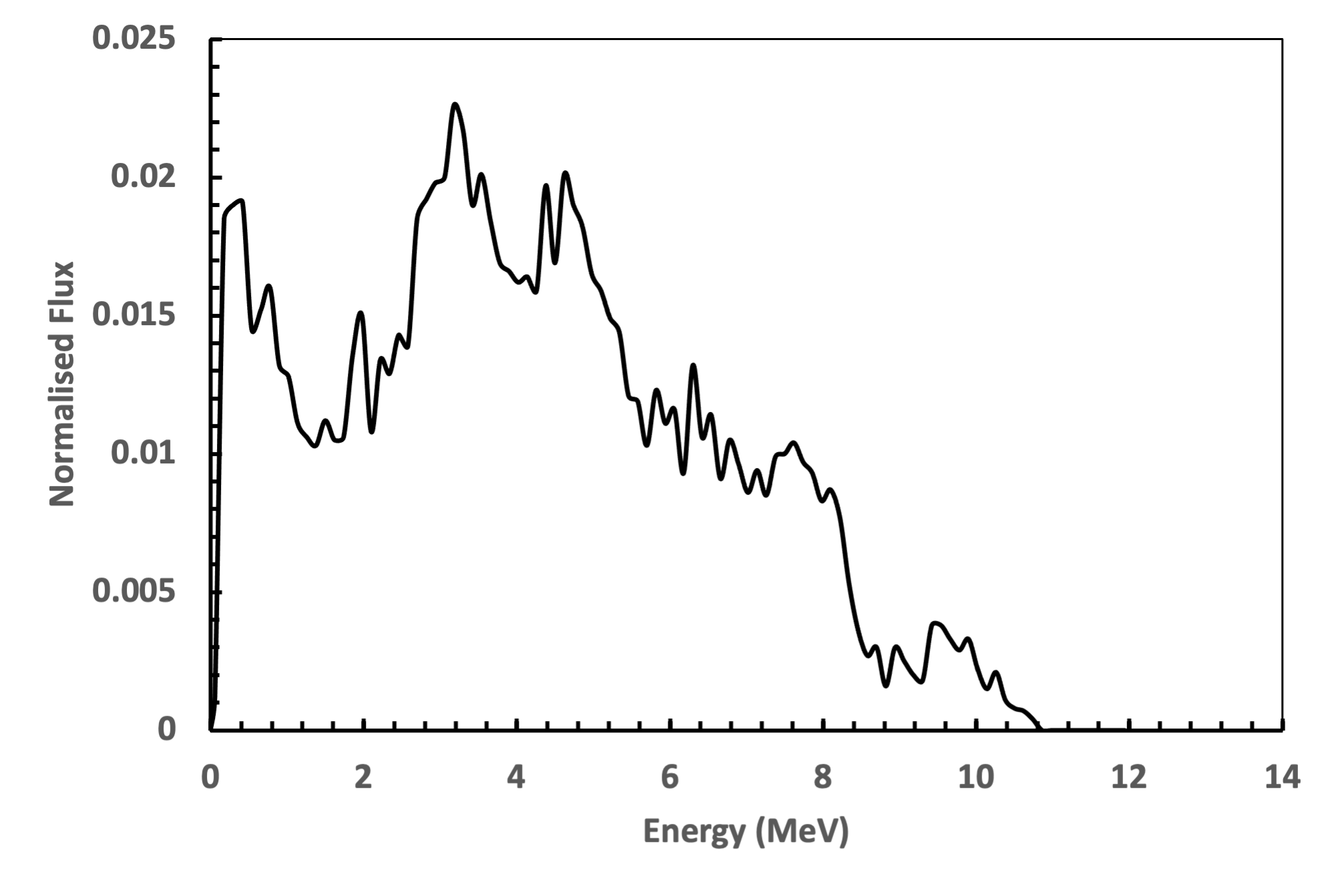}
 \caption{Modelled Am-Be source with GEANT4}
 \label{fig:Fig1}
\end{figure}

\section{Simulations}
\subsection{Modelling of detectors} 
The GEANT4 Monte Carlo simulation toolkit was originally developed for high energy physics community and recent developments and consistent expansion of physics models led to its application for thermal neutron energies \citep{Zugec2016,LoMeo2015}. To initiate the neutron events, and for specifying parameters like energy, direction, number of incident neutrons, we use "General Particle Source" (GPS) module \cite{GPS}. For the present studies, we modelled a source-detector geometry configuration comprising of a polyethylene moderated Am-Be source emitting fast neutrons isotropically. The detector modelled is a GAGG:Ce crystal with 10 mm radius and varying the thicknesses. The libraries required for simulating thermal neutron induced reactions are present in GEANT4 in the form of G4ENDL (Evaluated Nuclear Data Library).

\subsection{Modelling of Am-Be Source}
For modelling the Am-Be neutron source, we consider a cylinder with dimensions of 10 mm diameter and 30 mm length and Am-Be radioactive material with an outer 2 mm coating of stainless steel. G4RadioactiveDecayPhysics is used for  initiating the alpha ($\alpha$) particle generation from Americium-241. These $\alpha$-particles impinge on the Beryllium nuclei resulting in the production of fast neutrons. The normalized energy distribution of fast neutrons emitted from Am-Be source is shown in Fig \ref{fig:Fig1} and is found to be in good agreement with the energy distribution reported in the literature \cite{Kluge}. 

\begin{figure}[bt]
 \centering
 \includegraphics[height = 2.5in]{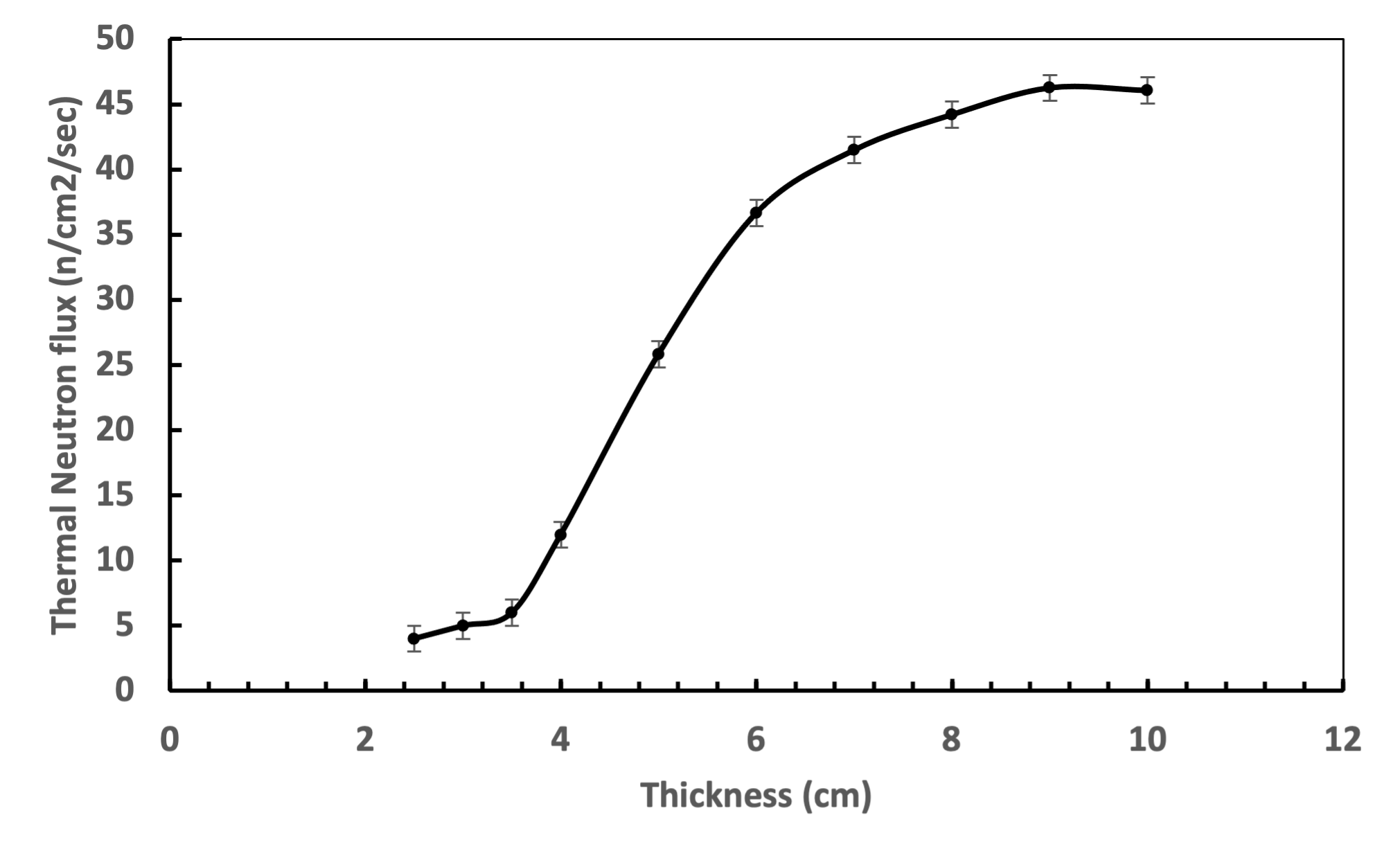}
 \caption{Plot of thermal neutron flux versus moderator thickness}
 \label{fig:Fig2}
\end{figure}

The fast neutrons so produced are then moderated using high density polyethylene (HDPE) sheets. The variation of thermal neutron flux with thickness of the moderator sheet is plotted in fig \ref{fig:Fig2}. It can be seen  that with increase in the moderator thickness, the flux of thermal neutrons increases and reaches maximum for a thickness of 9 cm which is in good agreement with the measurements reported in the literature \citep{Naqvi2006,Waheed2017}. 

\section{Experimental Setup}  
The experimental setup consists of a thermal neutron assembly comprising of a 500 mCi Am-Be source and suitable moderating assembly of HDPE sheets. A slice of 0.5 mm and 3 mm crystal is coupled to a PMT (model: ET9266) placed at distance of 5 cm from the source encapsulated in 9 cm of HDPE moderator to detect thermal neutrons. The scintillation decay times for gamma and neutron irradiation were measured directly from the anode of the PMT using a fast digital oscilloscope. A CAEN digital pulse processor along with the integrated MCA was used to process the signal from the PMT.

\section{Results and Discussion}
\subsection{Optimization of physics list in GEANT4} 

In order to optimise the physics list of GEANT4 for thermal neutron simulations, we considered a cylindrical GAGG:Ce scintillating crystal of 10 mm radius and 0.5 mm thickness and a thermalised Am-Be source. 
\begin{figure}[bt]
    \centering
    \includegraphics[height=2.5in]{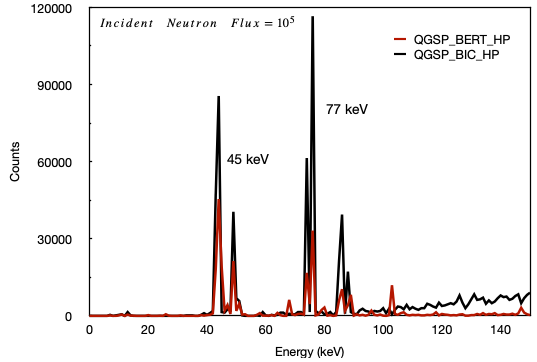}
    \caption{Simulated response of 0.5 mm thick GAGG:Ce detectors for thermal neutrons emitting from a thermalised Am-Be source for two physics lists.}
    \label{fig:Fig4}
\end{figure}
Based on the standard High Precision Neutron Package (Neutron\_HP)\cite{Allison2006}, there are two physics lists available in GEANT4 for neutron simulations namely, QGSP-BIC-HP  and QGSP-BERT-HP. While both the models use neutron cross sectional data from ENDF-B-VI library, the former uses binary cascading model \cite{Folger2004} where as the later uses Bertini cascading model\cite{Geng2014}. To choose the optimum physics list between the two, the simulations are made using both physics lists and the response is shown in figure \ref{fig:Fig4}. The spectra clearly show two peaks, around 45 keV and 77 keV, confirming the thermal neutron capture reaction with Gd isotopes as reported in\citep{Tyagi2019, Dumazert2018}. It is observed that the energy signatures for both the physics lists are the same but there is a clear difference in counts under the two peaks. As we are interested in knowing the peak areas only at this stage, the energy dependent resolution model is not applied for broadening the peaks due to which the peaks appears very narrow.
\begin{figure}[bt]
\centering
  \centering
  \includegraphics[height=2.5in]{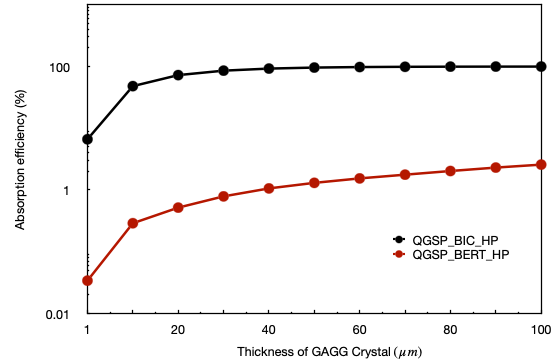}
  \caption{Thickness dependent thermal neutron absorption in GAGG:Ce for two physics lists in GEANT4}
  \label{fig:Fig5}
\end{figure}
To further investigate this, we plot the absorption efficiency (in percentage) of thermal neutrons in GAGG:Ce of different thicknesses considering both physics lists and results are shown in figure \ref{fig:Fig5}. Clearly, for "$QGSP\_BIC\_HP$", the total absorption of thermal neutrons is $100\%$ whereas it is around $22\%$ for "$QSGP\_BERT\_HP$". It is well known that due to very high thermal neutron capture cross section of $^{155}Gd$ and $^{157}Gd$ isotopes (61000 barns and 253700 barns respectively), the total absorption of thermal neutrons occurs in a detector thickness of about 0.1 mm only\cite{Reeder1994}. Thus, for all the simulations further done in this study, we have used $QGSP\_BIC\_HP$ which uses the neutron interaction cross sections from $ENDF/B-VI$. The low energy EM package is used to simulate the interaction of X-rays produced as a result of thermal neutron capture. It is to be noted that the thickness of about 0.1 mm is sufficient to fully absorb the thermal neutrons only but not the conversion electrons and X-rays produced as a result of thermal neutron capture. The total absorption of thermal neutron induced events requires more detector thickness which will be discussed later.


\subsection{Mean free path}

\begin{figure}[bt]
    \centering
    \includegraphics[height=2.5in]{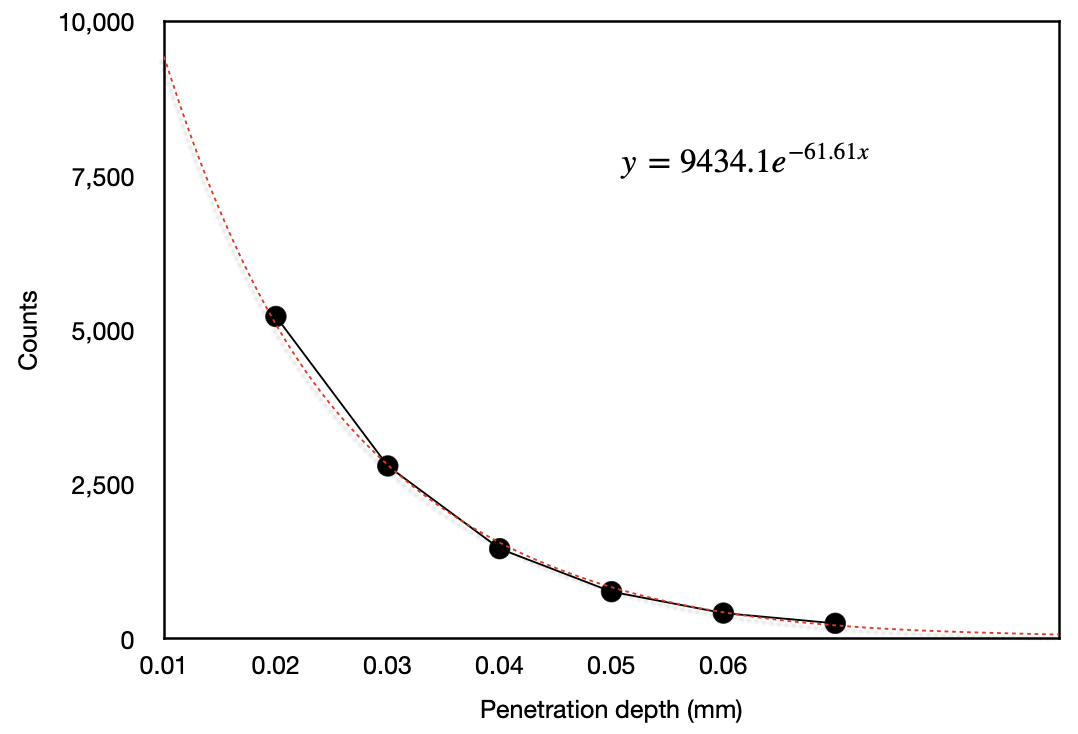}
    \caption{Effect of penetration depth of thermal neutrons on the neutrons emitted out of the GAGG:Ce crystal.}
    \label{fig:Fig6}
\end{figure}

As thermal neutrons get fully absorbed in about 0.1mm thick GAGG:Ce crystal, the interaction of thermal neutron inside a GAGG:Ce crystal can be better understood in terms of mean free path. In Gd based composite detector GSO, the mean free path has been reported to be 10 $\mu$m \cite{Reeder_1994}. In a recent study, GEANT4 simulation was performed on the penetrability of thermal neutrons in a cube of GAGG:Ce from which mean free path was determined to be 15.54 $\mu$m. \cite{Taggart2020}. In the present study, we performed GEANT4 simulations to understand the effect of penetration depth of the crystal on the number of thermal neutrons emitted out of the crystal and the results are shown in the Fig. \ref{fig:Fig6}.

\begin{table*}[!tbt]
    \caption{Sensitivity as a function of thickness of GAGG:Ce: Simulations and measurements.}
    \begin{tabular}{|c|cccc|c|c|}
    \hline
    \multirow{2}{*}{Thickness}& \multicolumn{4}{c|}{Simulations}&  \multicolumn{2}{c|}{Experiments}\\
    \cmidrule{2-7}
     (mm)& Area under & Area under & Area under & Sensitivity & Our Experi- & Taggart et.al\\
     & 32 keV& 45 keV & 77 keV & (77keV/45keV)  & -ments & (2020) \cite{Taggart2020} \\ \hline
        
         \hline
         0.01 & 36641& 4348 & 29769 &6.84(0.025) & & 6.921(0.016)\\
         0.05 &73255 & 11577 & 58354 &5.04(0.025) & & \\
         0.1 & 78729& 14674 & 60843 &4.146(0.03) & & 4.306(0.013)\\
         0.5 & 88848 & 24819 & 64473 &2.597(0.03)& 2.669 (0.006)&\\
         1 & 96158& 30632 & 67085 &2.19(0.03)&  & 2.219(0.005)\\
         1.5 &99636 & 34431 & 68658 &1.99(0.025)& &\\
         2 &103380 & 37101 & 70206 &1.89(0.025)& & \\
         2.5 &108106 & 39858 & 72563 &1.82(0.025)& & \\
         3.0 &110913 & 41941 & 74033 &1.76(0.025)& 1.731 (0.003) & \\
         3.5 & 113194 & 43826 & 74773 &1.70(0.03)&  &\\
         4 & 115673 & 45268 & 76201 &1.68(0.025)&  &\\
         4.5 & 117792 & 46705 & 77566 &1.66(0.025)&  &\\
         5 & 119740 & 48073 & 78669 &1.63(0.03)&  &\\
         6 & 124198 & 50599 & 81291 &1.60(0.025)&  &\\
         7 & 127045 & 52587 & 83222  &1.58(0.025)&  &\\
         8 & 127045 & 52587 & 83222 &1.58(0.025)&  &\\
         9 & 127045 & 52576 & 83222 &1.58(0.025)&  &\\
         10 & 127045 & 52587 & 83222 &1.58(0.025)&  &\\
         \hline
    \end{tabular}
    \label{tab:Table3}
\end{table*}

\begin{figure}[!tbt]
    \centering
    \includegraphics[height=2.5in]{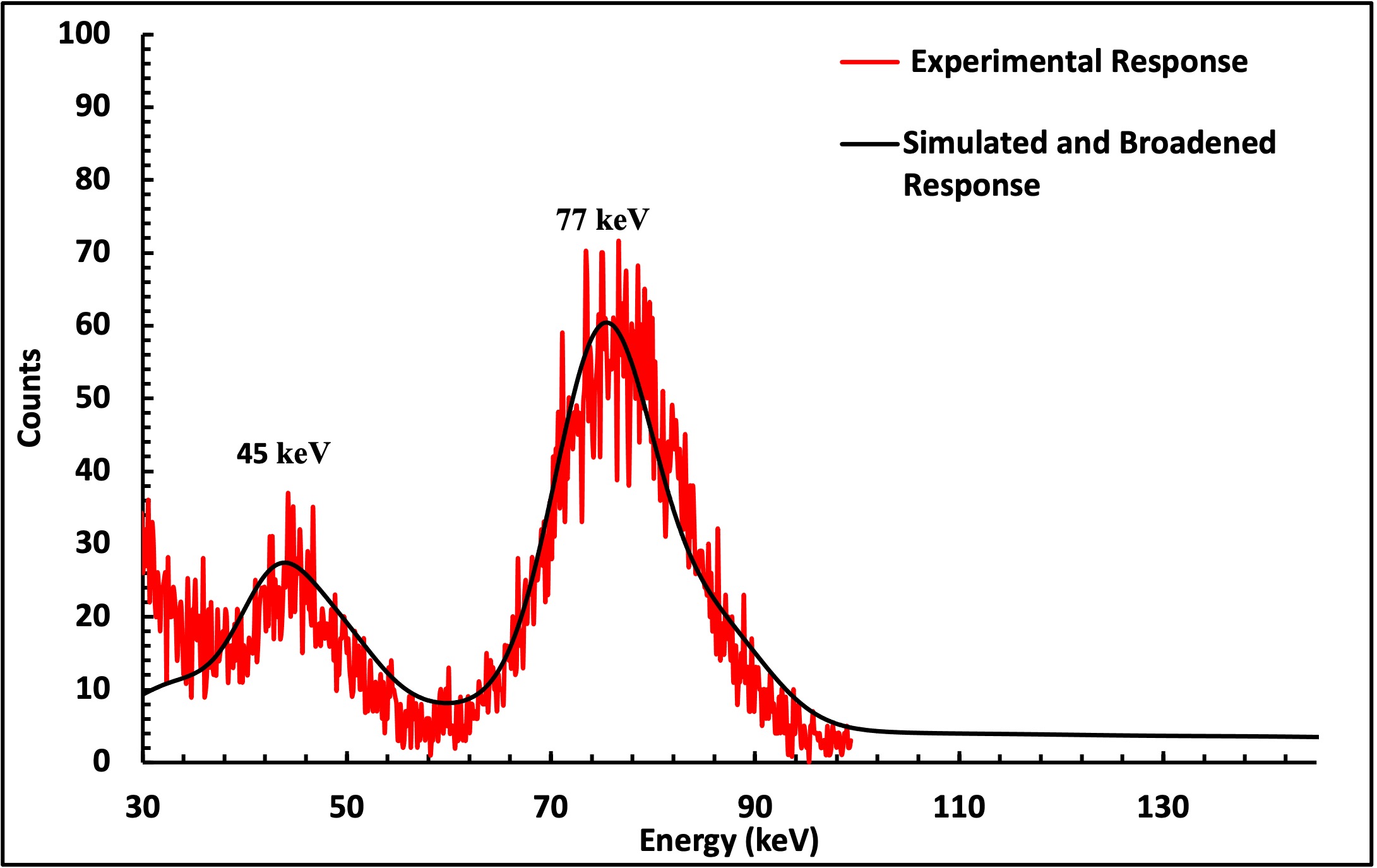}
    \caption{Response of 0.5 mm GAGG:Ce for thermal neutrons for moderated Am-Be}
    \label{fig:Spec1}
\end{figure}

\begin{figure}[!tbt]
    \centering
    \includegraphics[height=2.5in]{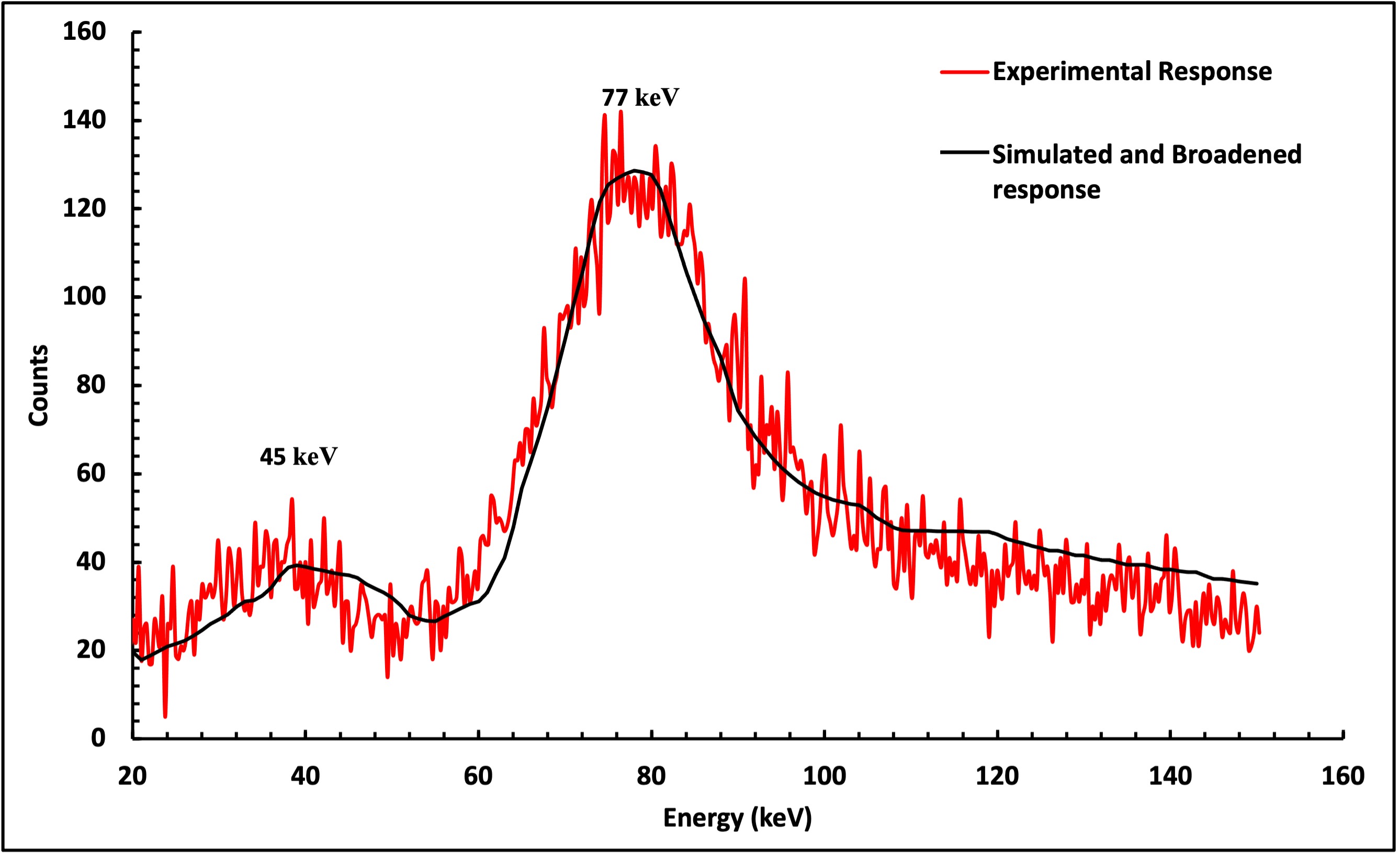}
    \caption{Response of 3 mm GAGG:Ce for thermal neutrons for moderated Am-Be}
    \label{fig:Spec2}
\end{figure}

We have calculated the macroscopic transport cross section of thermal neutron in GAGG:Ce to be $\Sigma_{tr}=61.61 cm^{-1}$ resulting in the mean free path $\lambda_{tr} = 16.23\pm 0.04 \mu m  \quad(\lambda_{tr}=\frac{1}{\Sigma_{tr}})$ which is approximately similar to the value reported in \cite{Taggart2020}.

\subsection{Validation of simulations with experimental measurements}
The experimental measurements were carried out using GAGG:Ce crystals of two thicknesses, namely, 0.5 mm and 3 mm coupled to PMT followed by standard pulse processing modules and data acquisition system. The measured spectra of 0.5 mm and 3 mm GAGG:Ce for thermalised Am-Be source are shown in Fig.\ref{fig:Spec1} and Fig.\ref{fig:Spec2} respectively. The figures also show broadened simulated spectra obtained using the general energy dependent resolution model where the energy resolution increases as $1/\sqrt{E}$, where E is the energy (in keV) of the X-rays and conversion electrons produced due to neutron interaction with GAGG:Ce crystal. Clearly, our simulations could successfully reproduce the measured spectra.

\subsection{Sensitivity and its dependence on crystal thickness}

\begin{figure}[!tbt]
  \centering
  \includegraphics[height=2.75in]{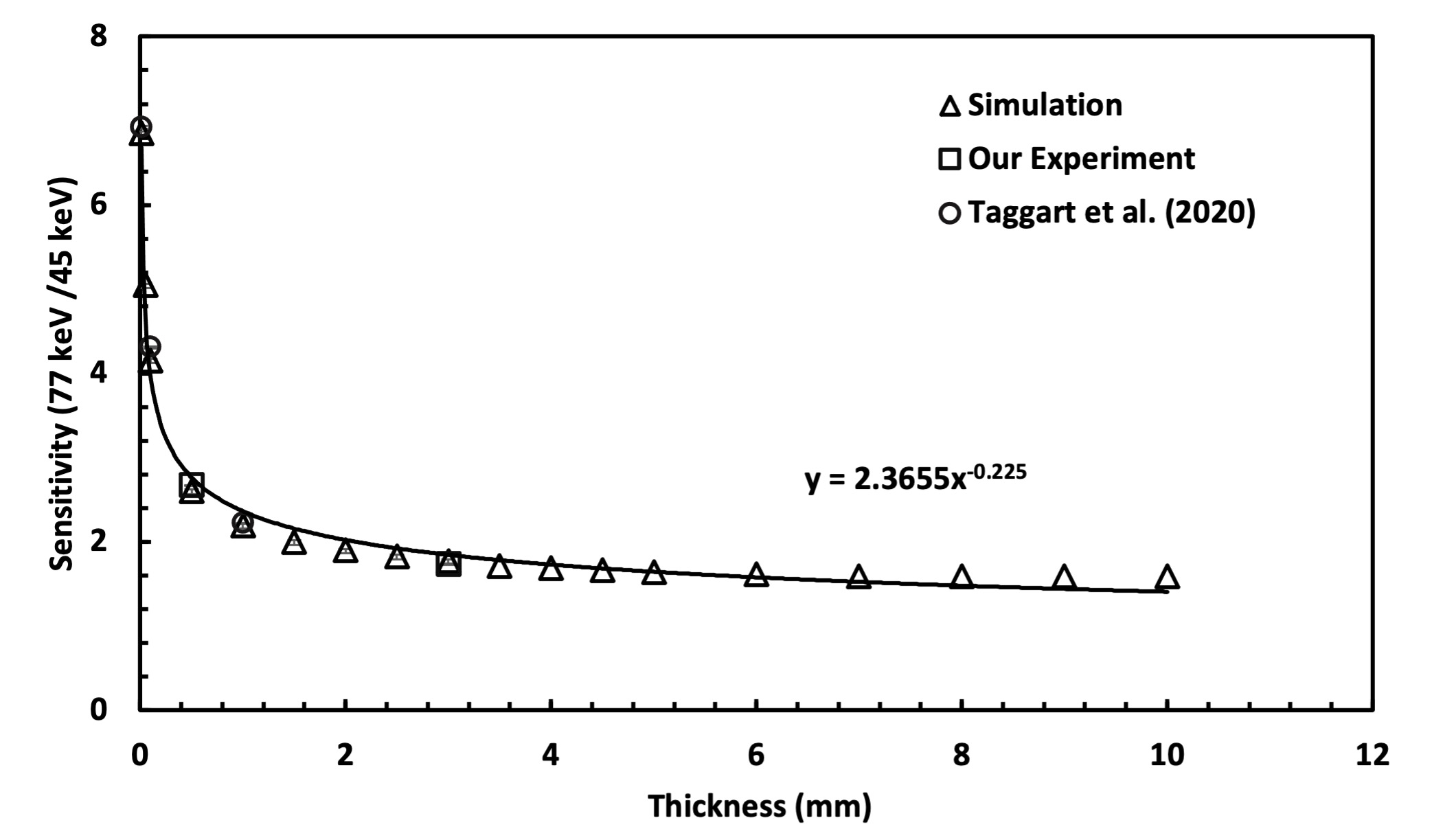}
  \caption{Plot of sensitivity vs thickness.}
  \label{fig:Fig9}
\end{figure}

In this study, we define the sensitivity of GAGG:Ce as the ratio of area under 77 keV sum peak to the area under the 45 keV peak. 
Simulations were done for various crystal thicknesses considering thermalised Am-Be source. The plot of sensitivity versus thicknesses is presented in Fig.\ref{fig:Fig9}. The sensitivity is found to decrease with increasing thickness of the GAGG:Ce crystal up to a value of 7 mm which means that the thermal neutron induced events (conversion electrons and X-rays) will get fully absorbed within 7 mm GAGG:Ce while thermal neutrons gets fully absorbed within 0.1 mm. 

 The results obtained from simulations and measurements in the present work are summarized in Table \ref{tab:Table3}. The table also shows the comparison between our simulations and measurements reported by Taggart et.al \cite{Taggart2020}. Overall, a good agreement between simulations and measurements is observed, thereby validating our simulations. The sensitivity can be calculated for a given thickness, up to 7 mm, by using an equation given by $y=2.3655x^{-0.225}$ where $'x'$ is the thickness in millimeter and $'y'$ is the sensitivity of GAGG:Ce detector. This data is important for selecting the crystals thickness for various applications. For example, if GAGG:Ce is to be chosen for imaging application, thicker crystals will provide lower spatial resolutions due to internal photon scattering effect, so the thinnest possible crystal that provides high sensitivity is preferred as previously discussed in \cite{Nikl2006,Kanai2013,Sen2017,Gerasymov2020}.

\section{Summary}
GAGG:Ce detector is proved to be one of the most attractive thermal neutron detectors in the recent times. In the present paper, we have proposed a general equation to estimate the sensitivity of GAGG:Ce detector based on the ratio of area under 77 keV sum peak to the area under 45 keV peak of thermal neutron induced spectrum. Extensive GEANT4 Monte Carlo simulations were made to understand the energy response of GAGG:Ce detector for thermal neutrons. Our simulations could successfully reproduce the entire measured spectra useful for estimating the sensitivity. The equation relating the sensitivity and crystal thickness follows an exponential pattern and the behaviour of sensitivity increases with decreasing thickness and is in good agreement with the measured results. The proposed equation is expected to be useful for manufacturers working in designing crystals for various applications like medical imaging, or cameras for nuclear security and safeguard devices in mixed radiation environment. 

\bibliography{mybibfile} 

\end{document}